\documentclass[namedreferences]{solarphysics}
\usepackage[optionalrh]{spr-sola-addons} % For Solar Physics 
\usepackage{graphicx}        % For eps figures, newer & more powerfull
\usepackage{color}           % For color text: \color command
\usepackage{url}             % For breaking URLs easily trough lines
\usepackage[pdfborder={0 0 0 },urlcolor=blue,breaklinks]{hyperref} 
\ifx \doiurl \undefined \def \doiurl#1{\href{http://dx.doi.org/#1}{\url{#1}}}\fi
\ifx \adsurl \undefined \def \adsurl#1{\href{http://adsabs.harvard.edu/abs/#1}{\url{#1}}}\fi

\newcommand{\aap}{    {\it Astron. Astrophys.}}

\newcommand{\apj}{    {\it Astrophys. J.}}
\newcommand{\apjl}{   {\it Astrophys. J. Lett.}}

\newcommand{\grl}{    {\it Geophys. Res. Lett.}}

\newcommand{\jgr}{    {\it J. Geophys. Res.}}

\newcommand{\solphys}{{\it Solar Phys.}}

\newcommand{\ssr}{    {\it Space Sci. Rev.}}

\begin{document}
\begin{article}
\begin{opening}

\title{Coronal Mass Ejections Associated with Slow Long Duration Flares}

%%%%%%%%%%%%%%%%%%%%%%%%%%%%%%%%%%%%%%%%%%%%%%%%%%%
%% Authors Names

\author{U.~B\c ak-St\c e\' slicka$^{1,}$$^{2}$\sep S.~Ko\l oma\'nski$^{1}$\sep T.~Mrozek$^{1,}$$^{3}$}

%%%%%%%%%%%%%%%%%%%%%%%%%%%%%%%%%%%%%%%%%%%%%%%%%%%
%% Runningheads
%
\runningauthor{U. B\c ak-St\c e\' slicka {\it et al.}}
\runningtitle{Coronal Mass Ejections Associated with Slow Long Duration Flares}

%%%%%%%%%%%%%%%%%%%%%%%%%%%%%%%%%%%%%%%%%%%%%%%%%%%
%% Affilations 
%
\institute{$^{1}$ Astronomical Institute, University of Wroc{\l}aw, ul. Kopernika 11, 51-622 Wroc{\l}aw, Poland
                     email: \url{bak,kolomanski,mrozek@astro.uni.wroc.pl}\\ 
             $^{2}$ Visiting Scientist, High Altitude Observatory, NCAR, P.O. Box 3000, Boulder, CO 80307, USA\\
             $^{3}$ Solar Physics Division Space Research Centre, Polish Academy of Sciences, ul. Kopernika 11, 51-622 Wroc{\l}aw, Poland}
%                     email: \url{e.mail-c} \\
%            }

%%%%%%%%%%%%%%%%%%%%%%%%%%%%%%%%%%%%%%%%%%%%%%%%%%%%%%%%%%%%%%%%%%%%%%%%
%%% Abstract 

\begin{abstract}
It is well known that there is temporal relationship between coronal mass ejections (CMEs) and associated flares. The duration of the acceleration phase is related to the duration of the rise phase of a flare. We investigate CMEs associated with slow long duration events (LDEs), \textit{i.e}. flares with the long rising phase. We determined the relationships between flares and CMEs and analyzed the CME kinematics in detail. The parameters of the flares (GOES flux, duration of the rising phase) show strong correlations with the CME parameters (velocity, acceleration during main acceleration phase and duration of the CME acceleration phase). These correlations confirm the strong relation between slow LDEs and CMEs. We also analyzed the relation between the parameters of the CMEs, \textit{i.e}. a velocity, an acceleration during the main acceleration phase, a duration of the acceleration phase, and a height of a CME at the end of the acceleration phase. 
The CMEs associated with the slow LDEs are characterized by high velocity during the propagation phase, with the median equal $1423$~km s$^{-1}$. In half of the analyzed cases, the main acceleration was low ($a<300$~m s$^{-2}$), which suggests that the high velocity is caused by the prolongated acceleration phase (the median for the duration of the acceleration phase is equal $90$ minutes). The CMEs were accelerated up to several solar radii (with the median $\approx 7\,R_{\odot}$), which is much higher than in typical impulsive CMEs. Therefore, slow LDEs may potentially precede extremely strong geomagnetic storms. The analysis of slow LDEs and associated CMEs may give important information for developing more accurate space weather forecasts, especially for extreme events.
\end{abstract}

%%%%%%%%%%%%%%%%%%%%%%%%%%%%%%%%%%%%%%%%%%%%%%%%%%%
%% Keywords
%
\keywords{Sun: coronal mass ejections (CMEs) - flares}
\end{opening}

%-------------------------------------------------------------------------------------------------------------------
%----------------------------------------  Introduction --------------------------------------------------------
%-------------------------------------------------------------------------------------------------------------------
\section{Introduction}
Coronal mass ejections (CMEs) are large-scale ejections of magnetized plasma from the solar corona. The first CME was clearly identified by \inlinecite{tousey1973} in the 1971 {\em Orbiting Solar Observatory-7} observations. Later, CMEs (previously named "coronal transient") were observed by {\em Skylab} \cite{gosling1974} and since that time they have been extensively studied. 

It is well known that CMEs are associated with other solar active phenomena such as solar flares and eruptive prominences \cite{munro1979}. \inlinecite{gosling1976} analyzed several dozen of events and found that for all events the average CME speed was $470$~km s$^{-1}$. The flare-associated CMEs were characterized by higher velocities with average value of 775~km s$^{-1}$. Events associated with eruptive prominences had average velocity 330~km s$^{-1}$. Based on observations of twelve events, \inlinecite{macqueen1983} showed that the flare-associated CMEs are characterized by higher velocities and show approximately constant speeds with height. CMEs associated with the prominences are accelerated over larger height range and attain lower velocities than flare-associated CMEs. Using observations from the \textit{Large Angle and Spectrometric Coronagraph} onboard the {\em Solar and Heliospheric Observatory} (SOHO/LASCO: \opencite{brueckner1995}),  \inlinecite{sheeley1999} separated CMEs into two classes: i) gradual CMEs, which formed when prominences rose up and their leading edges accelerated gradually to reach velocities in the range $400$\,--\,$600$~km s$^{-1}$; ii) impulsive CMEs, often associated with flares, which showed higher speed ($>750$~km s$^{-1}$) and clear evidence of deceleration. The existence of two classes of CMEs was also confirmed by other authors \cite{stcyr1999,andrews2001,moon2002}. Based on a sample of $95$ CME events, \inlinecite{bein2012} compared characteristics of CMEs associated with flares and with filament eruptions. They found that CMEs associated with flares are characterized by higher peak acceleration and shorter acceleration - phase duration.

There are also observations that do not confirm differences between the two types of CMEs \cite{dere1999,feynman2004}. Statistical analysis of a large number of events was presented by \inlinecite{vrsnak2005}. The authors analyzed 545 flare-associated CMEs and 104 non-flare CMEs and found that both data sets show quite similar characteristics: in both samples significant fraction of CMEs was accelerated or decelerated and both samples include a comparable ratio of fast and slow CMEs. They concluded that there is a continuum of events rather than two distinct (flare associated and non-flare associated) classes of CMEs. 

Relationships between flares and CMEs, \textit{e.g.} SXR flux \textit{vs.} CME velocity, duration of the flare rising phase \textit{vs.} duration of the acceleration phase, were investigated by many authors \cite{zhang2001,moon2003, burkepile2004,vrsnak2005, maricic2007,chen2009,berkebile2012}. \inlinecite{zhang2001} investigated the temporal relationship between CMEs and associated solar flares. They found that kinematic evolution of three of the four CMEs can be described by a three-phase scenario:\\
 i) the initiation phase, characterized by slow speed of the CME edge, which can last tens of minutes and occur before the onset of the flare.\\
 ii) the impulsive acceleration phase (described below).\\
 iii) the propagation phase during which the CME can accelerate, decelerate or can have constant speed. \\

The second, impulsive acceleration (or main acceleration), phase is often well synchronized with the rising phase of the associated soft X-ray flare \cite{zhang2001} and this relationship was confirmed by other authors \cite{gallagher2003,zhang2004,maricic2007,bein2012}. Analysis of the kinematic evolution of CMEs and associated flares during the impulsive phase shows a strong relationship between the CME acceleration process and the energy release in the associated flare (\citeauthor{temmer2008} \citeyear{temmer2008,temmer2010} and references therein). Typically, CMEs are accelerated over a short distance range \cite{alexander2002,temmer2010, bein2011}, but sometimes they accelerate up to heights of several solar radii; in some cases even as high as $7$ R$_{\odot}$ \cite{zhang2001,zhang2004,vrsnak2001,maricic2004}.

During the acceleration phase CME accelerations may vary from some ten up to some $1000$~m~s$^{-2}$ \cite{wood1999,zhang2001}. The main acceleration depends on the duration of the acceleration phase \cite{zhang2006,vrsnak2007, bein2011}. A fast CME ($v > 1000$~km~s$^{-1}$) can be strongly accelerated over a short time, or weakly accelerated over an extended time interval of several hours \cite{zhang2004,vrsnak2007}. For CMEs associated with flares, the duration of the acceleration phase is related to the duration of the rise phase of a flare. CMEs associated with flares of slow rising phase are accelerated slowly but for a long time.

Here, we investigate the kinematic evolution of CMEs associated with long duration flares of slow rising phase (slow LDEs). This group of flares is characterized not only by a long decay but also by long rising phase, which may last more than 30 minutes \cite{hudson2000,bak2005,bak2011}. \textit{Reuven Ramaty High Energy Solar Spectroscopic Imager} (RHESSI) analysis of several slow LDEs is presented in our previous article \cite{bak2011}. This analysis revealed that during those flares thermal energy is released very slowly and decreased slowly after reaching the maximum value. Nevertheless the amount of total released energy in a slow LDE is huge ($10^{31}-10^{32}$~erg). Thus we can ask if CMEs related to these flares are also very energetic (\textit{i.e.} fast). If the answer is positive, slow LDE flares may have significant impact on space weather although they do not look as spectacular and dangerous as other more impulsive and powerful flares. 

%-------------------------------------------------------------------------------------------------------------------
%----------------------------------------  Data Analysis --------------------------------------------------------
%-------------------------------------------------------------------------------------------------------------------
\section{Data Analysis}

\begin{table}[!t]
%\begin{center}
	\caption{Data for slow LDE flares associated with analyzed CMEs}
	\vspace{3mm}
	\label{flares}
	\begin{footnotesize}
		\begin{tabular}{ccccccc}
\hline
%\tiny{}&\tiny{}&\tiny{}&\tiny{}&\multicolumn{5}{c|}{\tiny{}}\\
&&Flare&Flare& Rise phase&GOES&\\
&Date&onset&peak&duration&class&NOAA\\
&&[UT]&[UT]&t$_{fl}$ [min]&&\\
\hline
1.&23 Feb. 1997&02:21&03:33&72&B7.2&8019\\
2.&14 Nov. 1997&09:05&10:38&93&C4.6&8108\\
3.&20 Apr. 1998&09:38&10:21&43&M1.4&$^{*}$\\
4.&16 Jun. 1998&18:03&18:42&39&M1.0&8232\\
5.&05 May 2000&15:02&16:21&79&M1.5&8970\\
6.&16 Oct. 2000&06:40&07:28&48&M2.5&9182\\
7.&02 Feb. 2001&19:09&20:37&88&C3.3&-\\
8.&01 Apr. 2001&10:55&12:17&82&M5.5&9415\\
9.&03 Apr. 2001&03:25&03:57&32&X1.2&9415\\
10.&05 Apr. 2001&08:37&09:22&45&M8.4&9393\\
11.&15 May 2001&17:52&19:40&108&C4.0&9461\\
12.&28 Dec. 2001&19:37&20:45&68&X3.4&9767\\
13.&10 Mar. 2002&22:21&23:25&64&M2.3&9871\\
14.&21 May 2002&23:14&00:30&76&C9.7&9948\\
15.&24 Oct. 2003&02:27&02:54&27&M7.6&10486\\
16.&18 Nov. 2003&09:23&10:11&48&M4.5&10508\\
17.&28 Jul. 2004&02:32&06:09&217&C4.4&10652\\
18.&29 Jul. 2004&11:42&13:04&82&C2.1&10652\\
19.&31 Jul. 2004&05:16&06:57&101&C8.4&10652\\
20.&13 Jul. 2005&14:01&14:49&48&M5.0&10786\\
21.&14 Jul. 2005&10:16&10:55&39&X1.2&10786\\
22.&27 Jul. 2005&04:33&05:02&29&M3.7&10792\\
23.&06 Sep. 2005&19:32&22:02&150&M1.4&10808\\
24.&16 Mar. 2011&17:52&20:34&162&C3.7&11169\\
\hline
\multicolumn{7}{@{} l @{}}{$^{*}$ active-region complex consisting of ARs 8198, 8195, 8194, 8200} \\

	\end{tabular}
	\end{footnotesize}
%\end{center}
\end{table} 
We used GOES light curves to search for the slow LDE flares (GOES class from B to X). Using {\em Yohkoh}/\textit{Soft X-ray Telescope} (SXT) and SOHO/\textit{Extreme UV Imaging Telescope} (EIT) data we selected only limb or near-the limb events ($|\lambda| > 60^{\circ}$) to reduce projection effects. For all selected slow LDEs we measured duration of the rising phase, using the Solar-Geophysical Data (SGD) catalogue (in a few cases we corrected those values looking at GOES light curves) and checked if flares were associated with the CMEs observed by SOHO/LASCO coronographs. Spatial and temporal correlation between flares and CME was required. Using {\em Yohkoh}/SXT and/or EIT images we carefully determined the position of the flare. Slow LDEs are usually observed in high arcades for many hours \cite{bak2005,bak2011}, so even for flares behind the limb we could determine their latitude and active region with satisfying accuracy. Using SXT and EIT data it was possible to observe the early signature of the CME eruptions (EIT observations) or in several cases also X-ray plasma ejections (SXT observations). A CME was considered as associated with a given slow LDE only when a CME central position angle (CPA) was in agreement with the flare position. In addition to the location agreement, most of the associated CMEs were observed within several minutes (up to 30) before or after flare start. For one very long evolving event, we used only C2 and C3 data the time difference between flare start and time of first detection of the CME was larger (up to 60 minutes). For detailed analysis we selected 24 events. To reduce uncertainty in CME-flare association we selected only events during which there was no other CME-flare activity. Basic information about the slow LDEs is given in Table \ref{flares}.

\begin{figure}[!ht]
\vspace{0.5cm}
\includegraphics[width=3.9cm,angle=90]{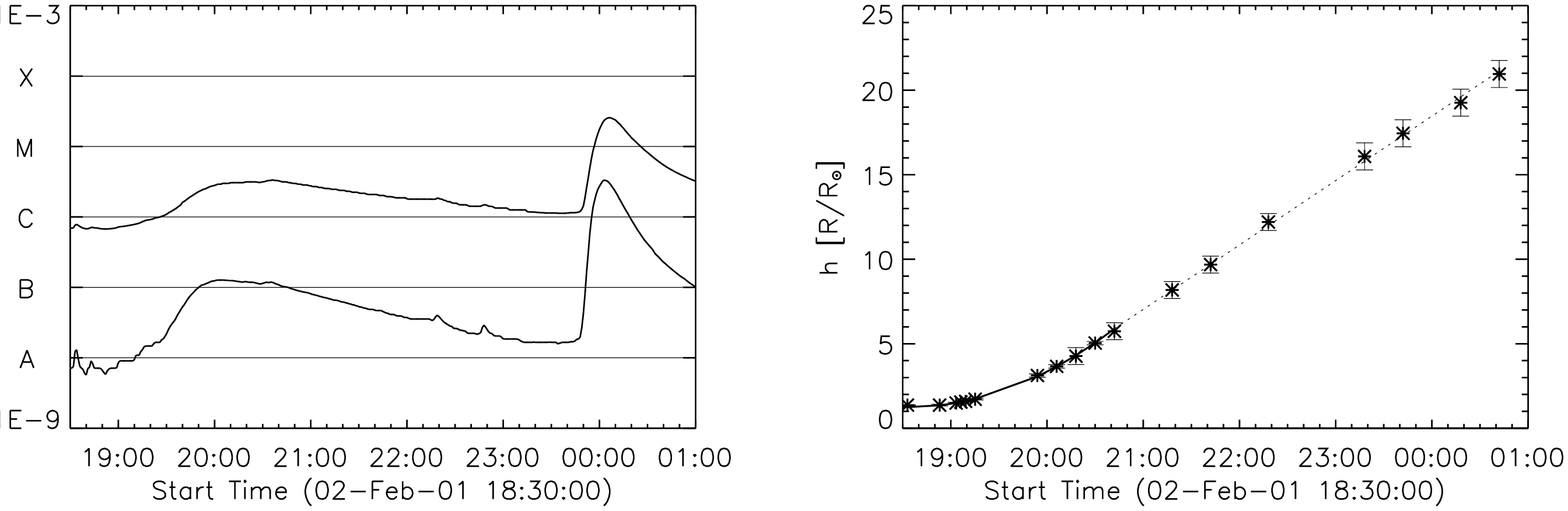}
\vspace{0.5cm}
\includegraphics[width=3.9cm,angle=90]{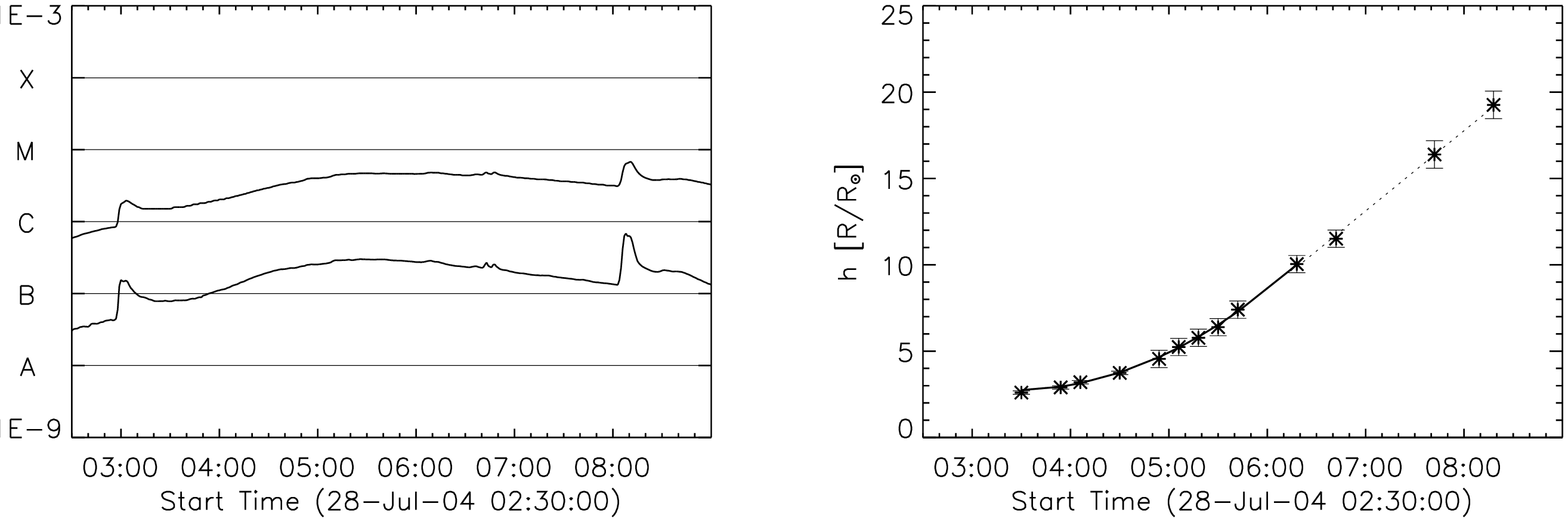}
\vspace{0.5cm}
\includegraphics[width=3.9cm,angle=90]{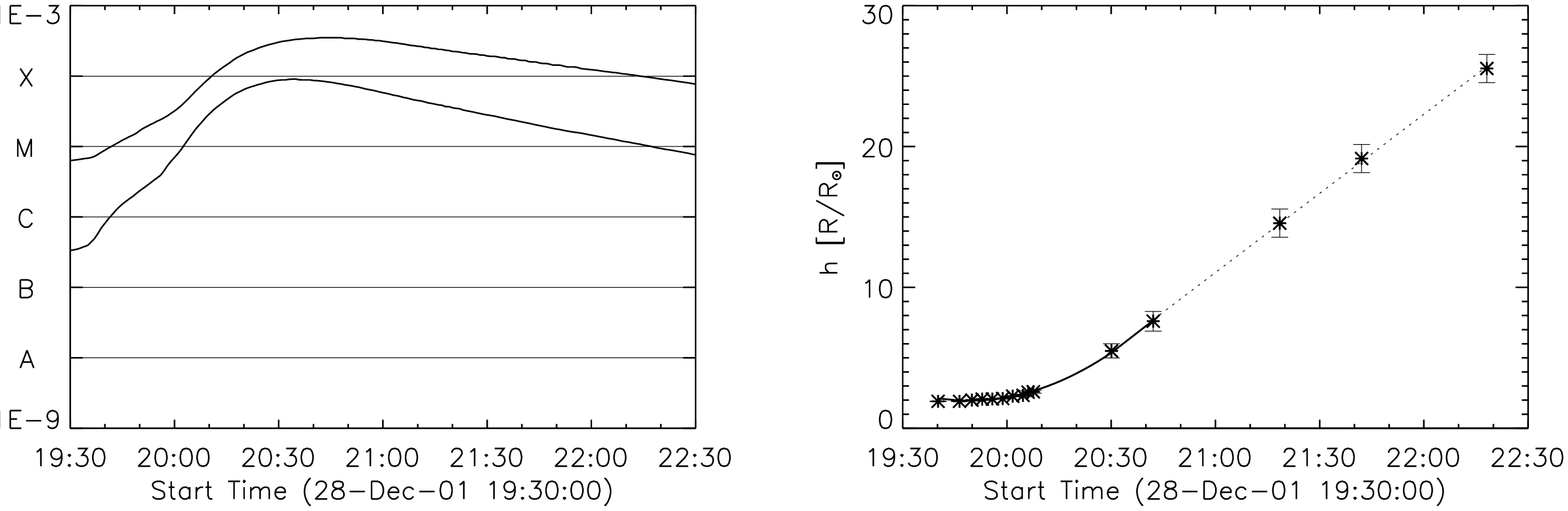}
\vspace{0.5cm}
\includegraphics[width=3.9cm,angle=90]{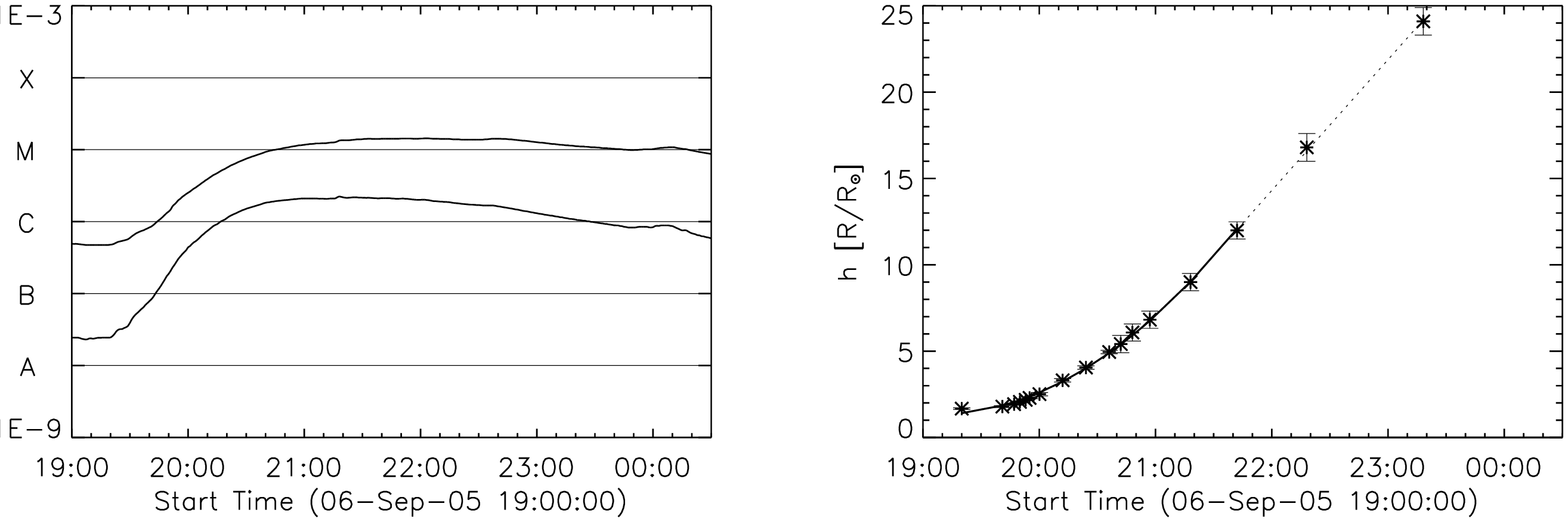}
\caption{Left: GOES X-ray fluxes of associated flares (upper curve: 1\,--\,8 \AA, lower curve: 0.5\,--\,4 \AA). Right: Height--time profiles of the CME's leading edge. The best fit to the height-time profile was with the second-degree polynomial (solid line) and straight line (dotted line).}
\label{sample_ht}
\end{figure}

Using running-differences images, we determined the heights of the leading edge of the CMEs in the plane of the sky. We visually followed the fastest segment of the CME leading edge and measured its height in several points; then we calculated an average value of $h$ that was used in further study. SOHO/LASCO-C2 and LASCO-C3 coronagraphs allow us to observe CMEs at heights 2.0\,--\,6.0\,$R_{\odot}$ and 3.7\,--\,30\,$R_{\odot}$ respectively \cite{brueckner1995}. CMEs associated with slow LDEs evolve slowly and are usually accelerated even in the LASCO-C2 field of view (see Figure \ref{sample_ht}); nevertheless to analyze the acceleration phase of the CME in more detailed, additional observations are needed. We used data from: the LASCO-C1 coronograph (which images the corona from $1.1$ to $3.0\,R_{\odot}$),  Mauna Loa Solar Observatory Mark-IV K-coronameter (MLSO/Mk4, at heights 1.14\,--\,2.86\,$R_{\odot}$), and SOHO/EIT (up to $1.5\,R_{\odot}$, \opencite{delaboudiniere1995}). In one case (No. 3) we also calculated heights of the radio type II burst source (AIP Potsdam observations, density model from \opencite{mann1999}) and assumed that this height is the same as that of the CME leading edge. Uncertainties of the CME heights may differ in various cases and depend on the sharpness of the leading edge of the CME. Based on previous reports \cite{tomczak2004,maricic2004,vrsnak2007} we assumed that the error of the measured height is: $0.05\,R_{\odot}$, $0.02\,R_{\odot}$, $0.01\,R_{\odot}$, $0.1\,R_{\odot}$, and 0.5\,--\,1.0\,$R_{\odot}$ for the MLSO/Mk4, EIT, LASCO-C1, LASCO-C2, and LASCO-C3 respectively.   

We fitted a combined function to the height--time profiles: a second-degree polynomial during acceleration phase and a linear fit during the propagation phase. The second-degree polynomial fit was applied for time $t_{\textrm {acc}}=t_{\textrm {tr}}-t_{\textrm {0}}$, where $t_{\textrm {0}}$ is the time of the beginning of the eruption or first CME observation. A linear fit was applied for $t=t_{\textrm {end}}-t_{\textrm {tr}}$, where $t_{\textrm {end}}$ is the time of the last observation. The transition point, $t_{\textrm {tr}}$ , was a free parameter calculated from the solution with the minimum standard deviation between the fitted curve and observed values of heights. This parameter gives us the end of the acceleration phase. Taking into account errors of the measured height and uncertainties of the fitting we obtained the uncertainty of the acceleration (up to $15$ \%) and uncertainty of the velocity (up to $6$ \%). The uncertainty of $t_{\textrm {acc}}$ depends on the temporal resolution of the observations and may change from few minutes to tens of minutes. In three cases (No. 17, 18, 19) acceleration was determined based on only C2 and C3 data (which was possible because of very slow evolution of the CMEs and very slow rising phase of associated flares), but this caused underestimation of $t_{\textrm {acc}}$. We also measured the height of the CME at the end of the acceleration phase [$h_{\textrm {acc}}=h(t_{\textrm {tr}})$]. GOES fluxes and CME height--time profiles (with fitted curves) for a few selected events are presented in Figure \ref{sample_ht}. 

%-------------------------------------------------------------------------------------------------------------------
%----------------------------------------  Results --------------------------------------------------------
%-------------------------------------------------------------------------------------------------------------------
\section{Results}
For the 24 CMEs associated with slow LDE flares we measured heights of the leading edge of CMEs. From the second-degree polynomial fitting we obtained: acceleration [$a$], during the main acceleration phase, duration of the acceleration phase [$t_{\textrm {acc}}=t_{\textrm {tr}}-t_{\textrm {0}}$], and the height at the end of the acceleration phase. The straight-line fitting for $t>t_{\textrm tr}$ gave us the average CME velocity [$v$], during the propagation phase. Parameters describing all analyzed CMEs are shown in Table \ref{CMEs}. Height--time profiles of the CMEs associated with the flares with the slowest rising phases show very slow evolution of the CMEs. These profiles are more similar to CMEs associated with the eruptive prominences, although they are associated with flares.  
 
\begin{table}[!t]
	\caption{Basic information about flares (duration of the rising phase [t$_{\textrm {fl}}$]; GOES class) and characteristics of analyzed CMEs (duration of the main acceleration phase [t$_{\textrm {acc}}$], average velocity during the propagation phase [$v$], mean acceleration during the acceleration phase [$a$], CME height at the end of the acceleration phase [$h_{\textrm {acc}}$], and list of the instruments used).}
	\vspace{3mm}
	\label{CMEs}
	\begin{footnotesize}
		\begin{tabular}{ccccccccc}
\hline
%\tiny{}&\tiny{}&\tiny{}&\tiny{}&\multicolumn{8}{c|}{\tiny{}}\\
&Date&GOES&t$_{\textrm {fl}}$&t$_{\textrm {acc}}$&$v$&$a$&$h_{\textrm {acc}}$&Instruments\\
&&class&\tiny{[min]}&\tiny{[min]}&\tiny{[km s$^{-1}$]}&\tiny{[m s$^{-2}$]}&\tiny{[$R_{\odot}$]}&\\
\hline
1.&23 Feb. 1997&B7.2&72&95&910&192&6.3&\tiny{EIT,C1,C2,C3}\\
2.&14 Nov. 1997&C4.6&93&122&1100&145&6.6&\tiny{C1,C2,C3}\\
3.&20 Apr. 1998&M1.4&43&67&1613&500&6.1&\tiny{EIT,radio,C2,C3}\\
4.&16 Jun. 1998&M1.0&39&39&1445&807&6.5&\tiny{EIT,Mk4,C1,C2,C3}\\
5.&05 May 2000&M1.5&79&78&1618&490&9.2&\tiny{EIT,C2,C3}\\
6.&16 Oct. 2000&M2.5&48&102&1340&245&7.1&\tiny{EIT,C2,C3}\\
7.&02 Feb. 2001&C3.3&88&109&731&96&5.7&\tiny{EIT,Mk4,C2,C3}\\
8.&01 Apr. 2001&M5.5&82&90&1524&205&9.7&\tiny{EIT,C2,C3}\\
9.&03 Apr. 2001&X1.2&32&78&1502&323&8.3&\tiny{EIT,C2,C3}\\
10.&05 Apr. 2001&M8.4&45&50&1769&561&5.9&\tiny{EIT,C2,C3}\\
11.&15 May 2001&C4.0&108&101&1227&281&6.8&\tiny{EIT,Mk4,C2,C3}\\
12.&28 Dec. 2001&X3.4&68&56&2173&785&7.6&\tiny{EIT,Mk4,C2,C3}\\
13.&10 Mar. 2002&M2.3&64&90&1403&302&9.1&\tiny{EIT,C2,C3}\\
14.&21 May 2002&C9.7&76&90&1256&282&6.9&\tiny{EIT,C2,C3}\\
15.&24 Oct. 2003&M7.6&27&54&1186&375&3.6&\tiny{EIT,C2,C3}\\
16.&18 Nov. 2003&M4.5&48&90&1881&379&9.6&\tiny{EIT,C2,C3}\\
17.&28 Jul. 2004&C4.4&217&168&903&99&10.0&\tiny{C2,C3}\\
18.&29 Jul. 2004&C2.1&82&108&1331&218&9.5&\tiny{C2,C3}\\
19.&31 Jul. 2004&C8.4&101&132&1304&231&11.9&\tiny{C2,C3}\\
20.&13 Jul. 2005&M5.0&48&66&1423&407&6.0&\tiny{EIT,C2,C3}\\
21.&14 Jul. 2005&X1.2&39&40&1663&965&5.9&\tiny{EIT,C2,C3}\\
22.&27 Jul. 2005&M3.7&29&54&1763&716&6.7&\tiny{EIT,C2,C3}\\
23.&06 Sep. 2005&M1.4&150&170&1455&172&12.0&\tiny{Mk4,C2,C3}\\
24.&16 Mar. 2011&C3.7&162&131&732&63&7.2&\tiny{Mk4,C2,C3}\\
\hline
	\end{tabular}
	\end{footnotesize}
\end{table}

All analyzed CMEs were associated with slow LDE flares, so we investigated the relationship between parameters describing flares (GOES flux, duration of the rising phase, $t_{\textrm {fl}}$) and CMEs (main acceleration, duration of the acceleration phase and velocity). For our sample of events we obtained:
\begin{itemize}
\item Velocity of the CMEs is well correlated with the peak of the soft X-ray (SXR) flux of the associated flare (correlation coefficient $r=0.77$). The fastest CMEs are associated with the strongest flares. This relation has been presented by other authors \cite{moon2003,burkepile2004, maricic2007,chen2009, bein2012} and is also observed in our sample of events (Figure \ref{goes_v_a}, left). It is worth noticing that CMEs connected with the slow LDEs are very fast. In $83 \%$ of cases the average velocity during the propagation phase was greater than $1000$~km s$^{-1}$, even if the associated flares were less powerful (GOES C class). All CMEs associated with M and X class flares have velocity greater than $1000$~km s$^{-1}$. The median is equal to $1423$~km s$^{-1}$. 
\item A similar relation can be observed between the peak of the soft X-ray flux and the acceleration of the CME during main acceleration phase (Figure \ref{goes_v_a}, right). Stronger acceleration was observed in the case of the CMEs associated with the stronger flares. This correlation is slightly weaker than the $v$ {\em vs.} GOES flux relation. The correlation coefficient is $0.66$.

\item CMEs parameters are also correlated with the duration of the rising phase of flares. CMEs with a larger value of the acceleration are associated with flares with a shorter rising phase. CMEs with the lowest value of the acceleration are associated with flares with the longest rising phase (Figure \ref{t_flare_a_tacc}, left). For our sample of events this relation is strong; the correlation coefficient is $-0.75$. 
 
\begin{figure}[!t]
\begin{center}
\includegraphics[width=6.03cm]{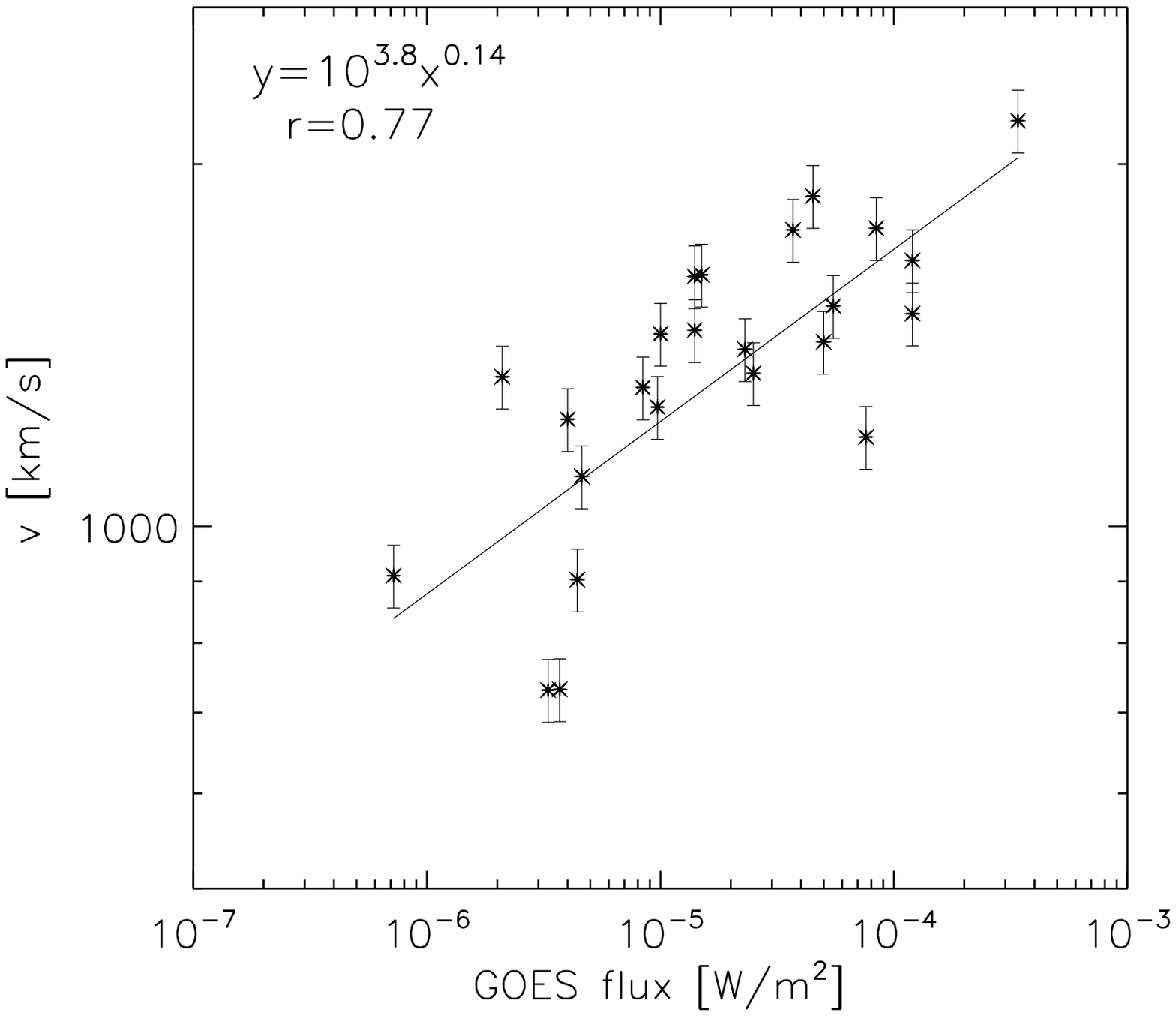}
\includegraphics[width=6.03cm]{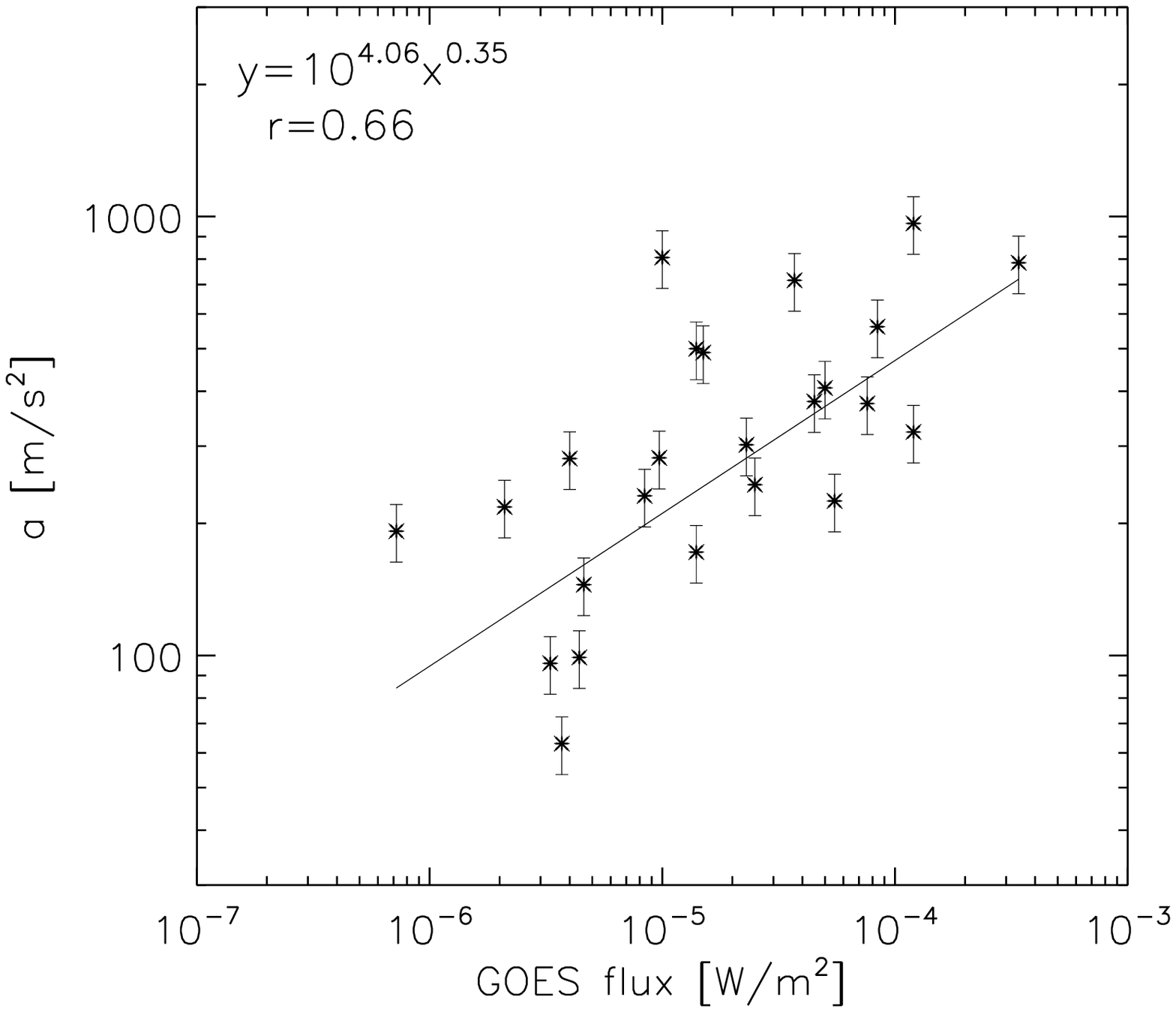}
\caption{Left: Relation between peak of the GOES X-ray flux (1\,--\,8~\AA) and CME velocity during the propagation phase. Right: Relation between peak of the GOES X-ray flux (1\,--\,8~\AA) and CMEs main acceleration (obtained from the quadratic fit). The straight lines show the best fit with a linear correlation coefficient [r].}
\label{goes_v_a}
\end{center}
\end{figure}

\begin{figure}[!t]
\begin{center}
\includegraphics[width=6.09cm]{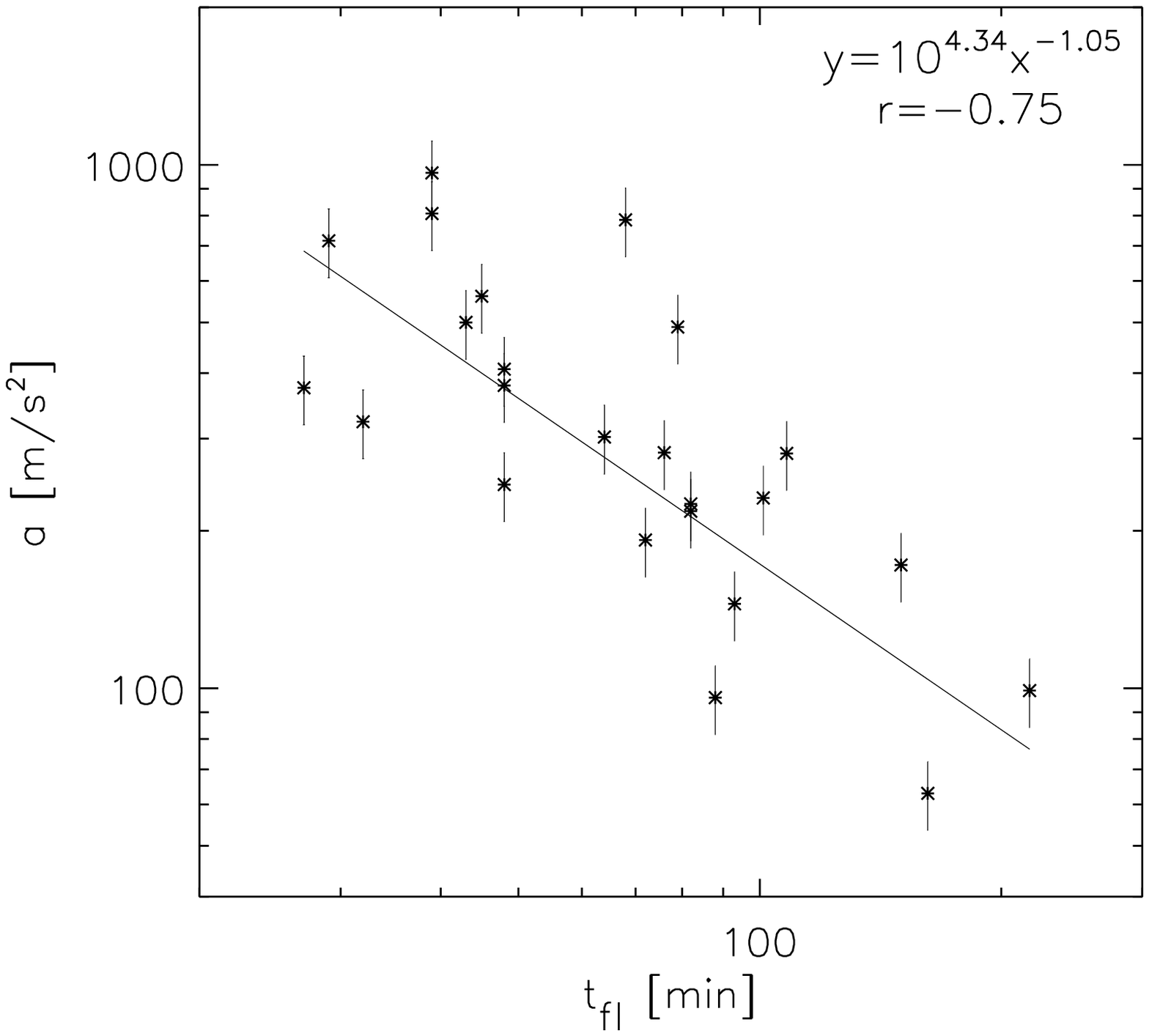}
\includegraphics[width=5.98cm]{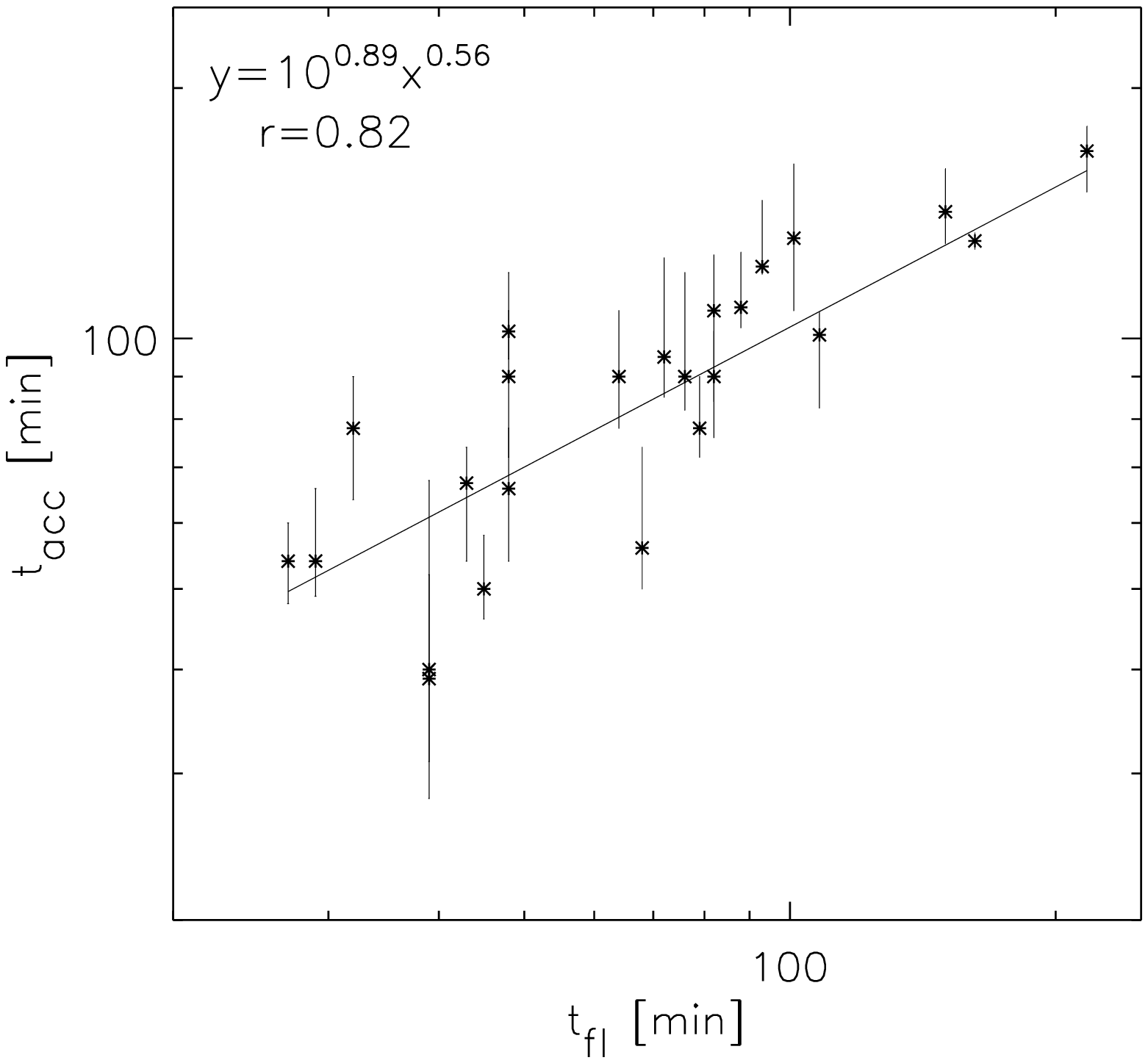}
\caption{Left: Relation between duration of flares rising phase and main acceleration (obtained from the quadratic fit) of the CMEs. Right: Relation between duration of rising phase of flares and duration of the CMEs acceleration phase. The straight lines show the best fit with the linear correlation coefficient [r].}
\label{t_flare_a_tacc}
\end{center}
\end{figure}

\item There is a strong correlation ($r=0.82$) between duration of the rising phase of flares and duration of CMEs acceleration phase (Figure \ref{t_flare_a_tacc}, right). CMEs with longer CME acceleration phase tend to be associated with flares with longer rising times (see also \opencite{maricic2007}). Median for $t_{\textrm {acc}}$ is equal to 90 minutes, and median for $t_{\textrm {fl}}$ is equal to 72 minutes.
\end{itemize}

All of these relations show a very strong connection between CMEs and slow LDE flares and suggest that those flares and CMEs are different manifestation of the same energy-release process. In the next step we have investigated relations between parameters describing CMEs.

\begin{figure}[!t]
\begin{center}
\includegraphics[width=6.03cm]{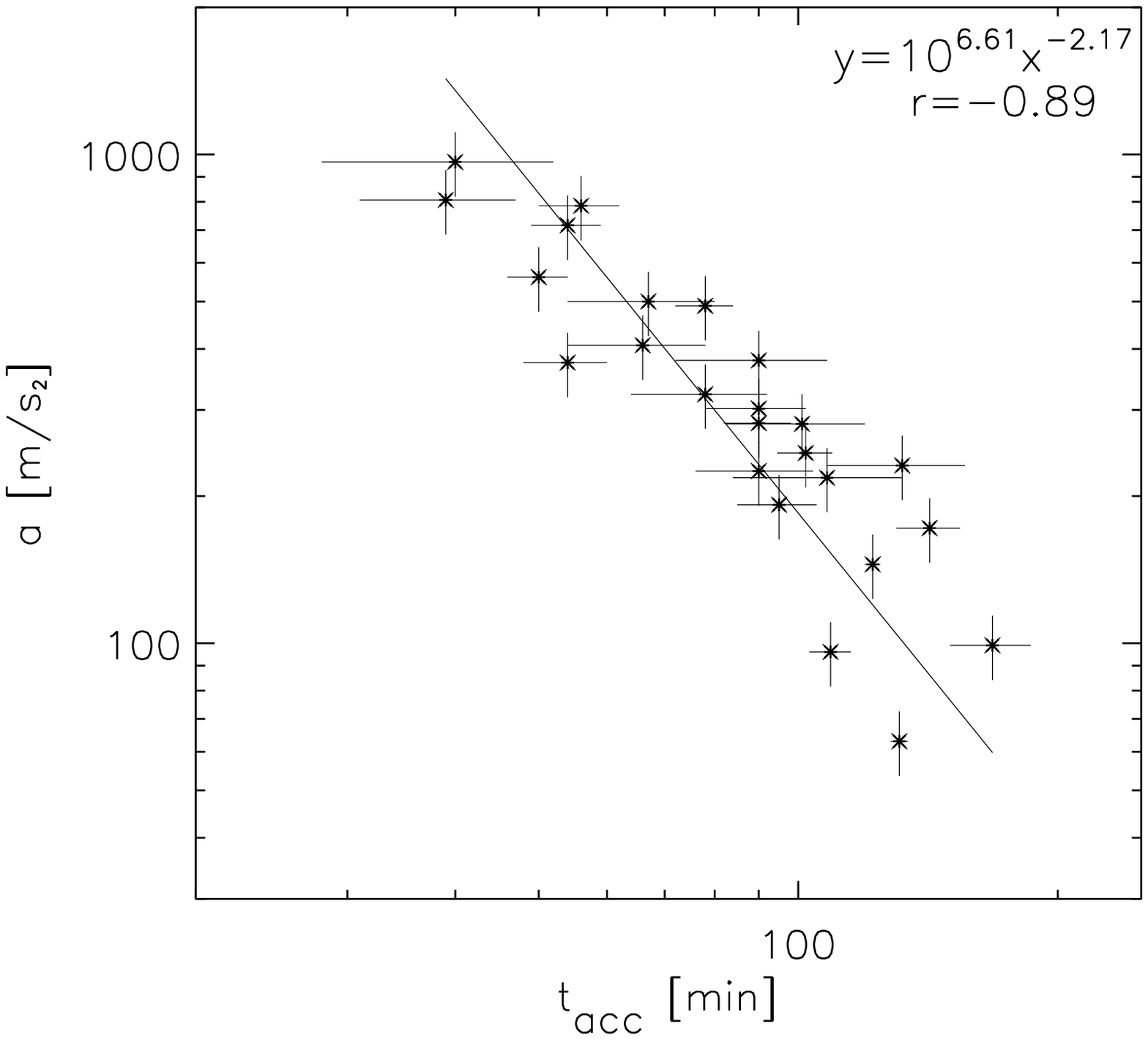}
\includegraphics[width=6.03cm]{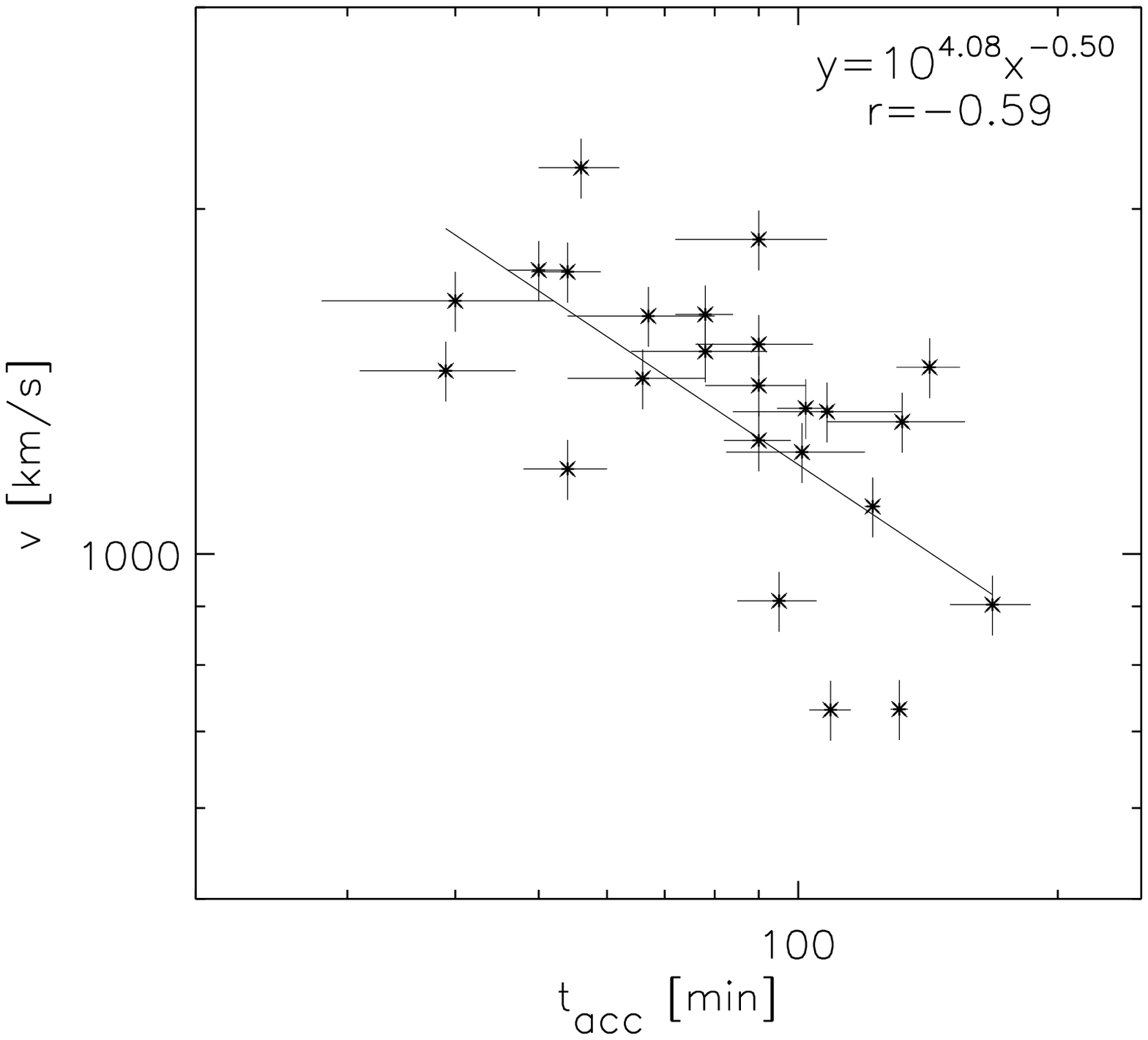}
\caption{Left: Relation between duration of the acceleration phase of CMEs and main acceleration (obtained from the quadratic fit) of the CMEs. Right: Relation between duration of the acceleration phase of CMEs and velocity of the CMEs during the propagation phase. The straight line show the best fit with a linear correlation coefficient [r].}
\label{tacc_a_v}
\end{center}
\end{figure}

\begin{figure}[!h]
\begin{center}
\includegraphics[width=5.90cm]{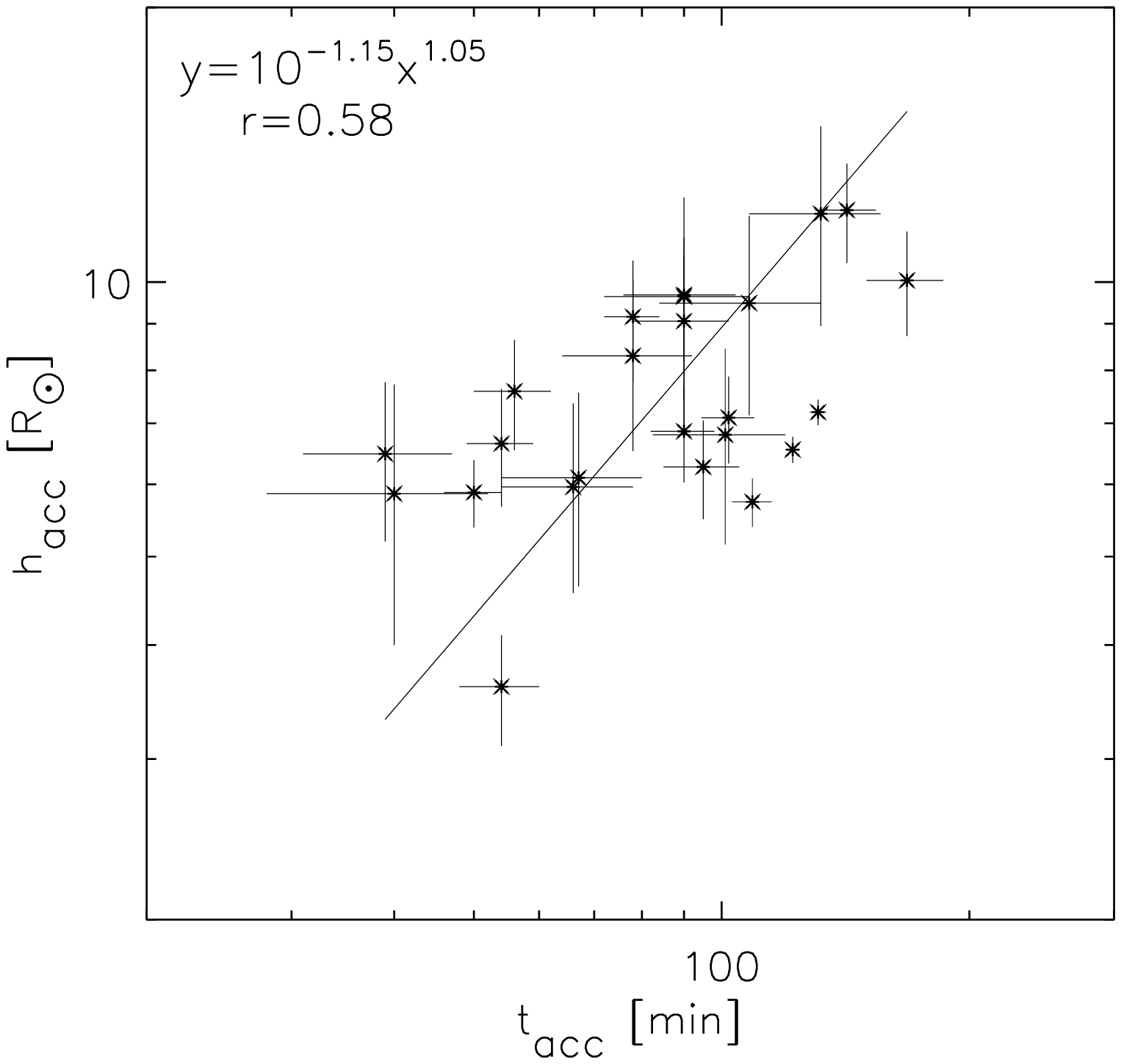}
\includegraphics[width=6.11cm]{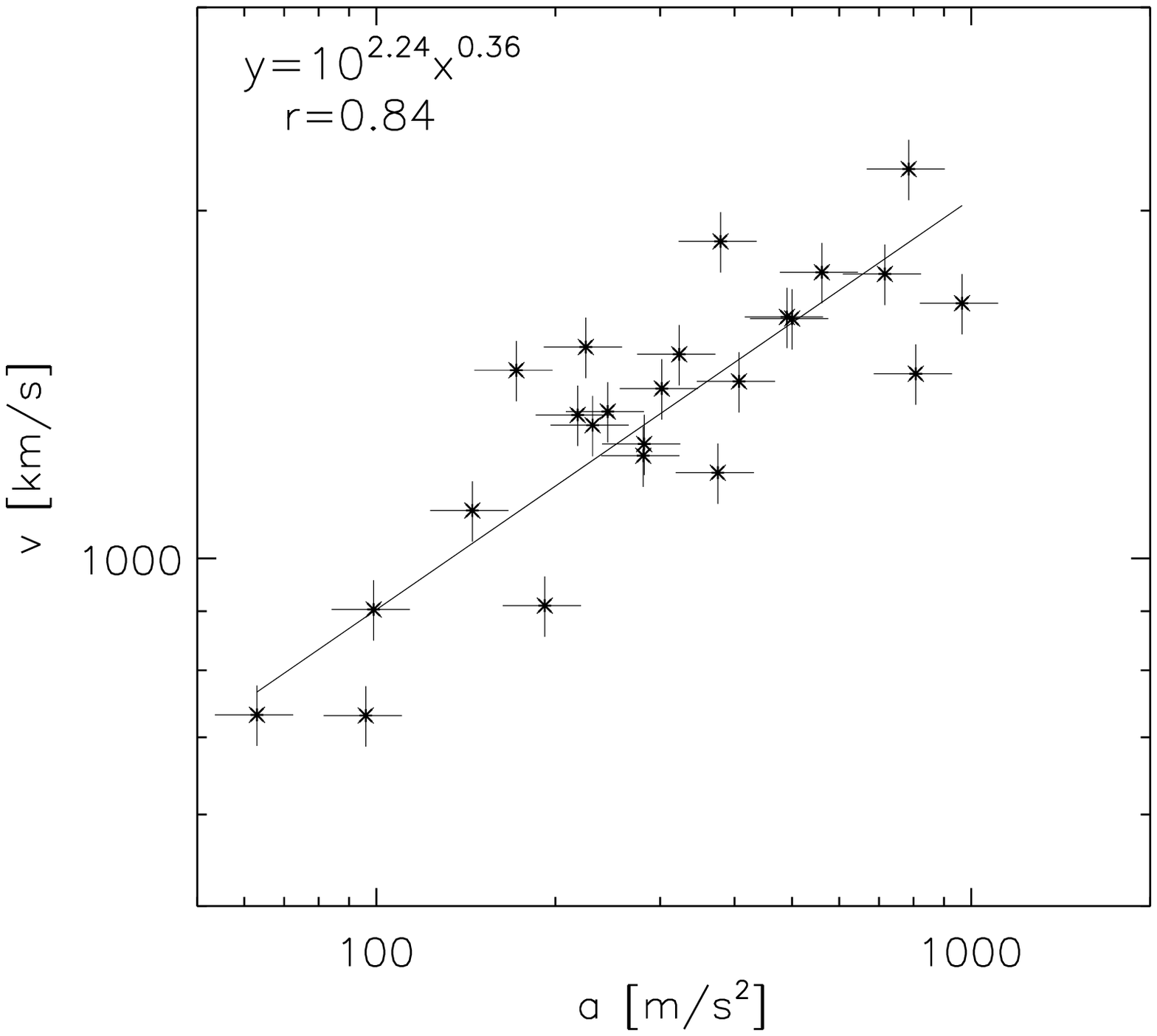}
\caption{Left: Relation between duration of the acceleration phase of CMEs and height of the CME at the end of the acceleration phase. Right: Relation between main acceleration (obtained from the quadratic fit) and velocity during the propagation phase. The straight line show the best fit with a correlation coefficient [r].}
\label{tacc_h_a_v}
\end{center}
\end{figure}
\begin{itemize}
\item Figure \ref{tacc_a_v} (left panel) shows strong correlation between duration of the acceleration phase and value of the main acceleration. CMEs accelerated for a longer time ($t_{\textrm {acc}}>90$~minutes) are characterized by low acceleration during the main acceleration phase ($a<300$~m~s$^{-2}$). The median value for $a$ is $302$~m~s$^{-2}$. This is the strongest relation that we found, with correlation coefficient $r=-0.89$. The relation between acceleration (peak or mean) and duration of the acceleration phase was presented earlier by other authors \cite{zhang2006,vrsnak2007,bein2011}.

\item There is also a relationship between duration of the acceleration phase and average velocity during the propagation phase (Figure \ref{tacc_a_v}, right). CMEs accelerated for a shorter time have larger velocity (and are accelerated more rapidly). CMEs accelerated for a longer time have lower velocity. However this relation varies slowly, thus even CMEs with a longer $t_{\textrm {acc}}$ have very high velocity (in most cases larger than $1000$~km s$^{-1}$).
\item Duration of the acceleration phase is also related to the height of CME at the end of accelerated phase [$h_{\textrm {acc}}$]. CMEs accelerated for longer time are accelerated to the greater height (Figure \ref{tacc_h_a_v}, left), in most cases $h_{\textrm {acc}}>5\,R_{\odot}$. The correlation coefficient is equal $0.58$, median for $h_{\textrm {acc}}$ is equal about $7\,R_{\odot}$. 
\item \inlinecite{bein2011} demonstrated a relation between CME peak acceleration and peak velocity for a large sample of events. For our sample of events we investigated the relationship between average velocity during the propagation phase and acceleration during the main acceleration phase (Figure \ref{tacc_h_a_v}, right). Correlation is strong with $r=0.84$. Stronger acceleration caused higher velocity during the propagation phase. In twelve cases, acceleration during the main acceleration phase was low ($a<300$~m~s$^{-2}$), despite of that the velocity was high ($v>1000$~km s$^{-1}$) in most cases. This means that this high velocity was caused by low, but prolonged, acceleration.
\end{itemize}
%-------------------------------------------------------------------------------------------------------------------
%----------------------------------------  Discussion and Conclusion------------------------------------------------
%-------------------------------------------------------------------------------------------------------------------
\section{Discussion and Conclusions}

Slow LDEs are very interesting events. Their rising phase is much longer than in typical flares, while their impulsive phase is weak or does not exist. In a previous article \cite{bak2011} we showed that despite the low value of the heating rate, during the whole rise phase of a slow LDE the total released energy is huge. In our examples presented by \inlinecite{bak2011} it is around $10^{31}-10^{32}$~erg. The same magnitude of energy is released during the decay phase of LDEs \cite{kolomanski2011}. This value is larger, by at least an order of magnitude, than the total energy released during the rise phase of a short-rise flares of comparable GOES magnitude. 

CMEs associated with such gradual flares evolve slowly. The acceleration phase corresponds to the rising phase of a flare and prolonged acceleration causes high CME velocity. We obtained a median value for CME velocity of $1423$~km s$^{-1}$, which is a much larger value than in all previous studies. Recently \inlinecite{bein2012} analyzed $70$ CMEs associated with flares. This is the largest sample of CMEs for which the acceleration phase was analyzed in detail. They found the median for $v_{max}$ equal to $461$~km s$^{-1}$, but their analysis was based on rather weak flares (only five M class flares, no X class). Even if we limit our sample to only B and C class flares, associated CMEs have median velocity of $1100$~km s$^{-1}$. Therefore higher velocity in our sample of events does not result from the presence of stronger flares. For our events with a large span in GOES class (from B to X class) we obtained a good correlation ($r=0.77$) between the SXR flux and CME velocity. This is a much stronger correlation than the relation presented by \inlinecite{bein2012}. It may suggest that this group of CMEs is very consistent and strongly associated with slow LDEs. 

A relation between CMEs and associated slow LDEs is also observed in the $a$ {\em vs.} GOES flux relation (Figure \ref{goes_v_a}, right). A similar relation was presented by other authors \cite{maricic2007,bein2012}, but they determined the peak of the CME acceleration, not the mean value of acceleration as in our cases. The power-law relation with the index $0.35$ obtained for our sample of events (with $r=0.66$) is similar to the relation obtained by \inlinecite{maricic2007} (power-law index $0.36$) but this differs from the relation obtained by \inlinecite{bein2012} (power-law index $0.14$). This difference may be caused (as in the previous relation) by the fact that the authors analyzed flares in a narrow GOES class range. \inlinecite{reeves2010} used a model of solar eruptions to investigate the relationship between the CME kinematics, thermal-energy release, and soft X-ray emissions in solar eruptions. They modeled a power-law relationship between the peak acceleration of a flux rope and the peak GOES flux with the indexes 0.39\,--\,0.49.

Not only the GOES flux correlates with the CME parameters. Duration of the rising phase of a flare is also correlated to the CME acceleration (Figure \ref{t_flare_a_tacc}, left) and duration of the acceleration phase (Figure \ref{t_flare_a_tacc}, right). This first correlation is quite strong ($r=-0.75$). The second relation ($t_{\textrm {fl}}$ {\em vs.} $t_{\textrm {acc}}$) is even stronger ($r=0.82$) with the power-law index $0.56$. The same positive correlation was presented by \inlinecite{maricic2007}, but the power-law index was different, $1.35$. This difference may be caused by the different method used to fit to the CME height--time profile. For most of our events the duration of the CME acceleration phase was longer than the duration of the flare rising phase, which is consistent with the results obtained by \inlinecite{maricic2007}. For our sample of events the median for $t_{\textrm {acc}}$ is 90 minutes and the median for $t_{\textrm {fl}}$ is 72 minutes.\\

Comparing to previous reports mentioned above, CMEs associated with slow LDEs showed stronger correlations between CMEs and flares parameters. This may be connected with the fact that we analyzed a narrowly selected group of events (\textit{i.e}. $t_{\textrm {fl}}>30$ minutes) or with an other yet unknown factor. All these relations show that even for gradual flares the duration of the CME acceleration phase corresponds to the duration of the flare rising phase. The second very important conclusion is that, despite low acceleration, CMEs associated with slow LDEs are much faster than the average CME due to their prolonged acceleration phase.\\

CME acceleration correlates with the duration of the acceleration phase. All CMEs with $t_{\textrm {acc}}>90$ minutes are characterized by low acceleration during the main acceleration phase ($a<300$~m~s$^{-2}$). This relation, for our sample of events, is steeper (power-law index $-2.17$) than the relation presented by other authors \cite{zhang2006,vrsnak2007,bein2011}. For the analysis made by \inlinecite{vrsnak2007} and \inlinecite{bein2011} it is respectively $-1.14$ and $-1.09$. In the last relation the authors also took into account CMEs without associated flares. The difference  in the inclination may be caused by the fact that we determined the mean CME acceleration during the main acceleration phase, which is smaller than the CME peak acceleration derived in other studies. Another reason for this difference may be the fact that in a few cases (with very long acceleration phases) we determined $t_{\textrm {acc}}$ based on only LASCO-C2 and LASCO-C3 data (see Table \ref{CMEs}), which may lead to underestimation of the $t_{\textrm {acc}}$. We may notice that if we exclude two CMEs with the lowest value of acceleration (events No. 7, 24) the relation becomes flatter than the current one and the result will be similar to results obtained by \inlinecite{vrsnak2007}. The correlation between $t_{\textrm {acc}}$ {\em vs.} $v$ is weaker ($r=-0.59$) than the previous one, but clearly visible. This relation is flat; thus even CMEs with the low value of acceleration have high velocity during the propagation phase (for most of the cases velocity is larger than $1000$~km s$^{-1}$). This high velocity is caused by the prolonged acceleration phase.  

There is another very interesting relation between duration of the acceleration phase and the height that the CME attains at the end of the acceleration phase. CMEs are usually accelerated in the low corona, but for our slowest evolving CMEs this height may reach even $10\,R_{\odot}$ (see Figures \ref{sample_ht}, \ref{tacc_h_a_v} left) and for most of the cases is was higher than $5\,R_{\odot}$. Statistical analysis \cite{bein2011} showed that a majority of CMEs were accelerated up to height $<1\,R_{\odot}$. For our sample of events the median value of CME height at the end of the acceleration phase is much higher, about $7\,R_{\odot}$. Such a prolonged acceleration phase during which CME attain very high height was reported earlier, \textit{e.g.} \inlinecite{maricic2004}. The CME analyzed by  \inlinecite{maricic2004} is a typical CME associated with a slow LDE and is on our list of events. In all previous reports the authors calculated the heights at which the CME reaches its maximum acceleration and maximum velocity, and it is hard to compare those values directly with our results. However, the height that we obtained for the end of the acceleration phase is much higher than previously reported.

The last interesting result is a correlation (with $r=0.84$) between acceleration during the main acceleration phase and average velocity during the propagation phase (Figure \ref{tacc_h_a_v}, right). A similar relation (between CME peak acceleration and peak velocity) was presented by \inlinecite{bein2011} for a large sample of events. The relation presented by \inlinecite{bein2011} is more scattered, but when we compare both plots we notice that our group lies at the upper end of the \inlinecite{bein2011} relation.

Slow LDEs are flares that are connected with very fast CMEs. Therefore, they are a group of events that may potentially precede strong geomagnetic storms. From this point of view the analysis of slow LDEs and associated CMEs may give important information for developing more accurate space-weather forecasts, especially for extreme events. The more detailed investigation of the conditions in the solar atmosphere that lead to the slow LDEs occurrence is needed. Moreover, the mechanism of the prolonged acceleration up to several solar radii should be carefully inspected from the observational and theoretical point of view. The work is in progress. In the next article we will investigate CMEs associated with slow LDEs in more detail: \textit{e.g.} estimation of the mass and energy of the CMEs and also analysis of acceleration process and configuration of magnetic field during such events.

%%%%%%%%%%%%%%%%%%%%%%%%%%%%%%

\begin{acks}
We thank Joan Burkepile, Jerzy Jakimiec, and Micha{\l} Tomczak for helpful comments and discussions. We also thank the anonymous referee for valuable comments, which improved the aricle. We acknowledge financial support from the Polish National Science Centre grants: 2011/03/B/ST9/00104 and 2011/01/M/ST9/06096. UBS also acknowledges financial support from the Human Capital Programme grant financed by the European Social Fund. We thank the SOHO, GOES, and MLSO teams for their open data policy. The MLSO/MK4 data was provided courtesy of the Mauna Loa Solar Observatory, operated by the High Altitude Observatory, as part of the National Center for Atmospheric Research (NCAR). NCAR is supported by the National Science Foundation.
\end{acks}

% Without BibTeX 

\end{article} 
\end{document}